\pdfoutput=1
\documentclass{article}
\usepackage[a4paper, margin=2.5cm]{geometry}
\usepackage{graphicx}
\usepackage{amsmath}
\usepackage{amssymb}
\usepackage{subcaption}
\usepackage[breaklinks=true]{hyperref}
\usepackage{slashed}
\usepackage{bbold}
\usepackage{braket}
\usepackage[title]{appendix}
\usepackage{authblk}
\usepackage{ragged2e}
\usepackage{ulem}
\usepackage{xcolor}
\usepackage[backend=biber,style=numeric,sorting=none]{biblatex}
\title{A perturbative construction of a modular $S_3$ model for quarks}
\author[ ]{Gabriele Battimelli \thanks{e-mail address: battimelli.gbrl@gmail.com}}

\author[1]{Davide Meloni \thanks{e-mail address: \texttt{davide.meloni@uniroma3.it}}}

\affil[1]{\small Dipartimento di Matematica e Fisica, Università di Roma Tre, via della Vasca Navale 84, 00146
Rome, Italy}

\date{}

\begin{document}

\maketitle

\begin{abstract} We present a modular-invariant supersymmetric model of flavour for quark masses and mixing, based on the group $\Gamma_2\cong S_3$. Hierarchies in the quark sector follow from the symmetric structure of the model, as they are generated by modular forms of different weights in the mass matrices, and no severe tuning of free superpotential coefficients is needed. Moreover, no flavons are employed besides the modulus $\tau$ which is, given the gCP symmetry imposed on the model, the only source of CP violation. 8 dimensionless observables (4 mass ratios and 4 CKM parameters) are reproduced by making use of 10 free model parameters (including real and imaginary part of the modulus) within the $1\sigma$ range making our model, to the best of our knowledge, the $S_3$ modular model for quarks with the fewest free parameters.
\end{abstract}

\tableofcontents

\clearpage

\section{Introduction}

The flavour structure of the Standard Model is one of the great puzzles of 
particle physics. The patterns of fermion masses and mixings, together with 
the origin of CP violation, remain unexplained and call for new physics 
beyond the established framework. Traditional approaches to this issue often 
imply the existence of a flavour symmetry constraining Yukawa interactions, 
which is spontaneously broken when one (or more) gauge singlet fields, called 
flavons, acquire a vacuum expectation value that minimizes a suitable scalar 
potential \cite{Feruglio_2015}, \cite{feruglio2021leptonflavoursymmetries}. 
A paradigmatic example is the $U(1)$ Froggatt-Nielsen mechanism 
\cite{FroggattNielsen_1979}, which gives rise to quark mass hierarchies 
through multiple flavon insertions in the Lagrangian terms. Models based on 
continuous and discrete, abelian and non-abelian symmetries have been 
proposed (see \cite{Feruglio_2015}, \cite{feruglio2021leptonflavoursymmetries}, \cite{altarellidiscretesym}, \cite{Dingsym}), but a unique solution has not yet emerged.\\
In \cite{feruglio2017neutrinomassesmodularforms} a novel approach inspired 
by string theory was proposed: modular invariance. In modular symmetric 
models (which are naturally implemented in supersymmetry) Yukawa couplings 
become modular forms, i.e.\ holomorphic functions of the modulus field $\tau$, 
transforming (as well as the matter fields) in irreducible representations of 
a finite modular group $\Gamma_N$. For $N\le5$, these modular groups are 
isomorphic to well-known discrete groups already employed in flavour model 
building ($S_3, A_4, S_4, A_5$); hence, modular symmetry provides a 
generalization of the discrete symmetry scheme, in which the overall 
transformation rule of the fields is non-linear. The theory is required to 
be invariant under the full action of modular transformations.\\
The modulus acquires a vev and, in minimal models with no flavons, this is 
the only source of flavour symmetry breaking; the mechanism behind this 
breaking is still an active research subject (see, for example, 
\cite{Novichkov2022}). In the limit of exact supersymmetry, the structure of 
mass matrices is fully determined and cannot receive further contributions, 
resulting in a limited number of free parameters; possible corrections can 
only arise from supersymmetry breaking or from a different choice of the 
K\"ahler potential of the theory. Moreover, combining modular invariance with 
a so-called generalized CP symmetry further constrains the model 
\cite{Novichkov_2019} and identifies the only source of CP violation in the 
real part of the vev acquired by the modulus $\tau$.\\
Several flavour models have been constructed following the modular symmetry 
bottom-up paradigm; the lepton and quark sectors have been treated 
separately, with a common modulus shared between them, and also in unified 
GUT descriptions. Some examples are constructions based on 
$\Gamma_3\cong A_4$ \cite{Chen_2021A4SU5}, \cite{Kobayashi_2018A4}, 
\cite{okada2020quarkleptonflavorscommon} and $\Gamma_4\cong S_4$ 
\cite{Qu_2021S4}, \cite{Wang_2021S4}. Higher levels, as well as the groups $\Gamma'_N$, have been 
employed as well (e.g.\ \cite{Ding_2019A5}, \cite{Liu_2019}).\\
In \cite{Novichkov_2021residual} it was pointed out that fermion mass 
hierarchies may arise due to the proximity of the modulus vev to a point of 
residual symmetry, i.e.\ a value of $\tau$ that does not break the modular 
symmetry completely. These ``fixed points'' are $\tau=i$, 
$\tau=\omega=e^{i\frac{2\pi}{3}}=-1/2+i\sqrt{3}/2$ and 
$\tau=i\infty$. With $S$ and $T$ denoting the finite modular group 
generators, the fixed points respectively preserve $Z_2^S$, $Z_3^{ST}$ and 
$Z_N^T$ symmetries\footnote{In the case of groups $\Gamma'_N$, the $Z^S_2$ 
residual symmetry is replaced by $Z^S_4$ and there is also a $Z^R_2$ 
symmetry, linked to the additional generator $R$, that is unbroken for any 
value of $\tau$.}. As regards quarks, this possibility has been investigated 
in \cite{petcov2023infinity} and \cite{Petcov_2023omega} for $\Gamma_3\cong 
A_4$ and in \cite{varzielas2023quarksmodulars4cusp} for $\Gamma_4\cong S_4$; 
however, in these constructions it is not straightforward to reproduce the CP 
phase of the CKM matrix unless multiple moduli are introduced or complex 
parameter values are adopted (so that gCP is explicitly broken and CP 
violation can no longer be exclusively attributed to the gCP-breaking vev of 
$\tau$).\\ 
In this work, we apply the principle of modular invariance and present a 
flavour model for the quark sector based on the group $\Gamma_2\cong S_3$ 
(the smallest non-abelian discrete group). In modular model building, $\Gamma_2\cong S_3$ has been used as a flavour symmetry for leptons 
\cite{Kobayashi_2018}, \cite{Meloni2023}, \cite{Marciano:2024nwm}, 
\cite{Okada_2019}, \cite{Vien:2026bjc} and in the context of GUTs \cite{Belfkir2025}, 
\cite{hernández2025}, \cite{Kobayashi_2020}. Non-GUT quark models based on $\Gamma_2\cong S_3$ have been 
discussed in \cite{feruglio2024} (where additional vector-like quarks are 
also present) and in \cite{Kobayashi_2019} and \cite{slidesMarrone} 
(that assume no additional matter fields beyond those of the MSSM 
and the modulus, and no additional symmetries beyond supersymmetry). As far 
as the number of independent parameters is concerned, in \cite{feruglio2024} 
(whose main purpose was to address the strong CP problem), 15 real 
dimensionless parameters (including $\text{Re}(\tau)$ and $\text{Im}(\tau)$) 
reproduce 8 dimensionless observables (4 quark mass ratios, 3 mixing angles 
and the CP phase). In \cite{Kobayashi_2019}, 18 parameters are used to fit 
the same 8 observables. Finally, in \cite{slidesMarrone} a model with 14 
real parameters in the quark mass matrices plus 2 from the modulus (shared 
with the lepton sector) is discussed; the lepton sector is described by 12 
real parameters, and 18 observables (6 quark masses, 4 CKM parameters, 3 
charged lepton masses, 2 neutrino mass differences and 3 neutrino mixing angles) 
are fitted. In all these models, hierarchies among superpotential coefficients 
are imposed by hand to reproduce the data. The aim of 
the present work is to investigate the possibility of building a minimal 
quark model with the following features:
\\
\begin{itemize}
    \item it introduces no additional matter fields besides the MSSM ones and 
    the modulus;
    \item it generates hierarchies without fine-tuning of 
superpotential coefficients. Hierarchical structures of mass matrices follow from symmetry assumptions alone, without the need to impose strong hierarchies among free parameters;
    \item it employs fewer free parameters in describing the quark sector than 
    existing models; in particular, by implementing a gCP symmetry, it 
    identifies the only source of CP violation in the vev acquired by 
    $\text{Re}(\tau)$ after spontaneous symmetry breaking;
    \item it admits a systematic perturbative expansion.
\end{itemize}
The paper is organized as follows: in Sect.\ref{overview} we briefly review the fundamental aspects of modular invariance and how to build a modular-invariant theory. In Sect.\ref{strategy}, we discuss a toy model with two generations in order to clarify our building strategy. In Sect.\ref{model}, we present our model and perform the perturbative analysis on the mass matrices. In Sect.\ref{numerical}, we explain the numerical procedure we followed in order to find best-fit values for model parameters and show our numerical results. In Sect.\ref{conclusions}, we draw our conclusions.

\section{Modular symmetry: an overview}
\label{overview}

\subsection{The modular group and modular forms}

We now present the bottom-up formalism outlined in \cite{feruglio2017neutrinomassesmodularforms} (and in \cite{Liu_2019}, for the groups $\Gamma'_N$). In the context of modular transformations we define the complex variable $\tau$, the modulus, which is restricted to the upper half of the complex plane $\mathcal{H}$: \begin{equation}
    \mathcal{H}=\{\tau\in \mathbb{C}|\ \text{Im}(\tau)>0\}.
\end{equation}
The modular group acts on $\tau$ through linear fractional transformations
\begin{equation} \label{eq:2}
    \tau \rightarrow \gamma\tau= \frac{a\tau+b}{c\tau+d},\quad (\gamma=\begin{bmatrix}
        a & b\\
        c & d
    \end{bmatrix},\ a,b,c,d\ \text{integers},\ ad-bc=1)
\end{equation}
and in a supersymmetric theory transformations are induced on the matter supermultiplets according to representations of $\Gamma_N$, the finite modular groups. By making use of holomorphic functions of the modulus $\tau$, the so-called modular forms of level $N$, which also transform following representations of $\Gamma_N$, it is possible to build supersymmetric modular-invariant theories for fermion masses where modular forms play the role of Yukawa couplings. In the particular cases where modular forms are constant functions, these constructions coincide with models dealing with discrete flavour symmetries (for small $N$, finite modular groups are isomorphic to permutation groups). A different paradigm arises when forms with a non-trivial dependence on $\tau$ are considered.\\
The special linear group
\begin{equation}
    \Gamma \equiv SL(2,\mathbb Z)
    = \left\{
    \begin{pmatrix} a & b \\ c & d \end{pmatrix}
    \,\middle|\,
    a,b,c,d\in\mathbb Z,\quad ad-bc=1
    \right\}
\end{equation}
acts on the upper half-plane $\mathcal H$ by fractional linear
transformations,
$\tau\mapsto\gamma\tau=(a\tau+b)/(c\tau+d)$.
Since $\gamma$ and $-\gamma$ induce the same transformation, the
modular group is
$\overline{\Gamma}=PSL(2,\mathbb Z)=\Gamma/\{\pm\mathbb1\}$.
The matrices
\begin{equation}
    S=\begin{pmatrix}0&1\\-1&0\end{pmatrix},
    \qquad
    T=\begin{pmatrix}1&1\\0&1\end{pmatrix}
\end{equation}
generate $\Gamma$ and act as $S:\tau\mapsto-1/\tau$ and
$T:\tau\mapsto\tau+1$. They satisfy
$S^2=-\mathbb1$ and $(ST)^3=\mathbb1$; their images in
$\overline{\Gamma}$ therefore obey
$S^2=(ST)^3=\mathbb1$.\\
The principal congruence subgroup of level $N$ is
\begin{equation}
    \Gamma(N)=
    \left\{\gamma\in\Gamma\mid
    \gamma\equiv\mathbb1\pmod N\right\}.
\end{equation}
It is an infinite normal subgroup of finite index in $\Gamma$,
with $\Gamma(1)=\Gamma$. Its image under the natural projection
$\pi:\Gamma\to\overline{\Gamma}$ is
\begin{equation}
    \overline{\Gamma}(N):=\pi(\Gamma(N))
    \simeq
    \Gamma(N)\big/
    \bigl(\Gamma(N)\cap\{\pm\mathbb1\}\bigr).
\end{equation}
For $N=1,2$, this amounts to quotienting by $\{\pm\mathbb1\}$,
whereas for $N>2$ the projection is injective because
$-\mathbb1\notin\Gamma(N)$.\\
The finite modular groups are defined by
$\Gamma_N=\overline{\Gamma}/\overline{\Gamma}(N)$.
Their generators satisfy
\begin{equation}
    S^2=(ST)^3=T^N=\mathbb1.
\end{equation}
These relations give a complete presentation for $2\leq N\leq5$;
for $N\geq6$, additional relations are required.
One may also define the homogeneous finite modular groups
$\Gamma'_N=\Gamma/\Gamma(N)$, which are double covers of
$\Gamma_N$ for $N>2$ \cite{Liu_2019}.
The central element $R$, represented by $-\mathbb1$, satisfies
\begin{equation}
    S^2=R,\qquad (ST)^3=T^N=R^2=\mathbb1,
    \qquad RT=TR.
\end{equation}
For $N=2$, however, $R=\mathbb1$ and
$\Gamma'_2\simeq\Gamma_2\simeq S_3$.\\
A modular form of level $N$ and integer weight $k$ is a function
$Y(\tau)$ holomorphic in $\mathcal H$ and at all cusps of
$\Gamma(N)$, satisfying
\begin{equation}
    Y(\gamma\tau)=(c\tau+d)^k\,Y(\tau),
    \qquad
    \gamma=\begin{pmatrix}a&b\\c&d\end{pmatrix}\in\Gamma(N).
\end{equation}
There are no nonzero modular forms of negative weight, and
weight-zero forms are constants. For $N=1,2$, applying this
transformation law to $-\mathbb1\in\Gamma(N)$ gives
$Y(\tau)=(-1)^k\,Y(\tau)$, so nonzero forms necessarily have even
weight. For $N>2$, odd weights are also allowed
\cite{Liu_2019}.\\
We henceforth restrict to $N=2$ and write the weight as $2k$,
with $k\geq0$. The modular forms of weight $2k$ constitute a
complex vector space $\mathcal M_{2k}(\Gamma(2))$ of dimension
$k+1$. A basis can be chosen such that
\begin{equation}
    Y_i(\gamma\tau)
    =(c\tau+d)^{2k}\rho([\gamma])_{ij}Y_j(\tau),
    \qquad \gamma\in\Gamma,
\end{equation}
where $[\gamma]$ denotes the image of $\gamma$ in $\Gamma_2$
and $\rho$ is a generally reducible unitary representation of
$\Gamma_2\simeq S_3$
\cite{feruglio2017neutrinomassesmodularforms}.
Thus, after factoring out $(c\tau+d)^{2k}$, the transformation
depends only on the corresponding finite-group element.\\
Multiplication adds modular weights, so these spaces form the
 ring
\begin{equation}
    \mathcal M(\Gamma(2))
    =\bigoplus_{k=0}^{\infty}\mathcal M_{2k}(\Gamma(2)).
\end{equation}
This ring is generated by two independent weight-two forms,
which transform as an $S_3$ doublet. Every form of weight $2k$
is a homogeneous polynomial of degree $k$ in these generators;
the resulting products can be organised into irreducible
$S_3$ multiplets using the tensor-product decomposition rules.

\subsection{Constructing a modular-invariant supersymmetric theory}

We briefly review how to deal with a modular-invariant theory. We will work in $N=1$ unbroken global supersymmetry. The action takes the general form:
\begin{equation}
    \mathcal{S}=\int d^4x\,d^2\theta\, d^2\overline{\theta}\:K(\Phi,\overline{\Phi})+\int d^4x\,d^2\theta\: W(\Phi)+h.c.
\end{equation}
$K(\Phi,\overline{\Phi})$ is the K\"ahler potential, a real gauge-invariant function of the chiral superfields $\Phi$ and their conjugates, and $W(\Phi)$, the superpotential, is a holomorphic gauge-invariant function of the chiral superfields $\Phi$. With $\Phi\equiv(\tau,\varphi)$ we denote collectively all chiral supermultiplets of the theory: the modulus $\tau$ and chiral supermultiplets $\varphi$ separated into sectors $\varphi^{(I)}$ (which contain, for example, $SU(2)_L$ quark doublets, singlets...). We focus on Yukawa interactions only and neglect gauge fields. We ask for invariance under transformations of the modular group $\Gamma$, under which the supermultiplets $\varphi^{(I)}$ of each sector $I$ are assumed to transform in a representation $\rho^{(I)}$ of a quotient group $\Gamma_N$:
\begin{equation}\label{eq:trasfrules}
    \begin{cases}
        \tau \rightarrow \frac{a\tau+b}{c\tau+d}\\
        \varphi^{(I)} \rightarrow (c\tau+d)^{-k_I}\rho^{(I)}\varphi^{(I)}
    \end{cases}
\end{equation}
We stress here that the supermultiplets $\varphi^{(I)}$ are not modular forms and there are no restrictions on the possible value of $k_I$: hence, following \cite{Meloni2023}, we will call them modular charges to distinguish them from modular weights. The invariance of the action $\mathcal{S}$ under (\ref{eq:trasfrules}) requires the invariance of the superpotential $W(\Phi)$ and the invariance of the K\"ahler potential up to a K\"ahler transformation:
\begin{equation}
    \begin{cases}
        W(\Phi)\rightarrow W(\Phi)\\
        K(\Phi,\overline{\Phi})\rightarrow K(\Phi,\overline{\Phi})+f(\Phi)+f(\overline{\Phi})
    \end{cases}.
\end{equation}
For the K\"ahler potential, we assume here a minimal form, invariant under (\ref{eq:trasfrules}) up to K\"ahler transformations. This is a standard choice in modular model building:
\begin{equation}\label{eq:kahlerpotential}
    K(\Phi,\overline{\Phi})=-\Lambda^2_\tau\,\text{ln}(-i\tau+i\overline{\tau})+\sum_I(-i\tau+i\overline{\tau})^{-k_I}|\varphi^{(I)}|^2,
\end{equation}
where $\Lambda_\tau$ has the dimension of a mass, since the modulus $\tau$ describes a dimensionless chiral supermultiplet. For our results, only the value of the vev of $\tau$ will be relevant. In the following we will simply use $\tau$ to refer to the vev of the modulus, too.\\
The invariance of the superpotential $W(\Phi)$ under modular transformations provides a strong restriction on the theory. We can write $W(\Phi)$ as a product of supermultiplets $\varphi^{(I_i)}$ coupled to modular forms $Y_{I_1...I_n}(\tau)$:
\begin{equation}\label{eq:superpotential}
    W(\Phi)=\sum g_{I_1...I_n}(Y_{I_1...I_n}(\tau)\varphi^{(I_1)}...\varphi^{(I_n)})_{\textbf{1}}.
\end{equation}
The sum is over all possible combinations of superfields (respecting gauge symmetries) contracted with $Y_{I_1...I_n}(\tau)$ to form independent singlets $\textbf{1}$.
In order to ensure modular invariance, the modular forms $Y_{I_1...I_n}(\tau)$ of weight $k_Y$ transform in the representation $\rho$ of $\Gamma_N$ as:
\begin{equation}
    Y_{I_1...I_n}(\tau)\rightarrow(c\tau+d)^{k_Y}\rho \,Y_{I_1...I_n}(\tau)
\end{equation}
with $k_Y$ and $\rho$ such that:
\begin{itemize}
    \item the weight $k_Y$ should compensate the overall charge of the product $\varphi^{(I_1)}...\varphi^{(I_n)}$:
    \begin{equation}
        k_Y=k_{I_1}+...+k_{I_n},
    \end{equation}
    \item the product $\rho\times\rho^{I_1}\times...\times\rho^{I_n}$ contains an invariant singlet.\\
\end{itemize}
In addition to modular symmetry, we will impose a gCP symmetry on our model. This kind of symmetry can be combined with flavour ones \cite{FerugliogCP}, \cite{HolthausengCP}; the formalism allowing to combine modular and gCP symmetries is presented in \cite{Novichkov_2019}. In our case (i.e. adopting the basis in which the generators of $\Gamma_2\cong S_3$ are represented by symmetric matrices, see the next section and Appendix \ref{app:s3} for details), imposing this symmetry will make all superpotential coefficients real numbers, while the normalization constant of modular forms a purely imaginary number.

\subsection{The \texorpdfstring{$\Gamma_2 \cong S_3$}{Gamma2 = S3} group}
$S_3$ is the group of all permutations of three objects. Some of its features are summarized in Appendix \ref{app:s3} and further details on group theoretic aspects can be found in \cite{Ishimori_2010}. It is characterized by three irreducible representations: a singlet $\textbf{1}$, a pseudo-singlet $\textbf{1}'$ and a doublet $\textbf{2}$, and it has two independent lowest-weighted modular forms of weight $k=2$, which will be denoted $Y_1(\tau)$ and $Y_2(\tau)$. In \cite{Kobayashi_2018}, these modular forms have been calculated in the real-symmetric basis of $S_3$ (generators expressed in this basis are reported in Appendix \ref{app:s3}).
$Y_1(\tau)$, $Y_2(\tau)$ have been built starting from the Dedekind eta function
\begin{equation}\label{eq:etadedekind}
    \eta(\tau)\equiv q^{\frac{1}{24}}\prod_{n=1}^{\infty}(1-q^n),\quad q\equiv e^{2\pi i\tau}
\end{equation}
and their expressions are:
\begin{equation}\label{eq:modularforms}
\begin{cases}
    Y_1(\tau)=\frac{c}{2}[\frac{\eta'(\frac{\tau}{2})}{\eta(\frac{\tau}{2})}+\frac{\eta'(\frac{\tau+1}{2})}{\eta(\frac{\tau+1}{2})}-\frac{8\eta'(2\tau)}{\eta(2\tau)}]\\
    Y_2(\tau)=\frac{c}{2}\sqrt{3}\cdot[\frac{\eta'(\frac{\tau}{2})}{\eta(\frac{\tau}{2})}-\frac{\eta'(\frac{\tau+1}{2})}{\eta(\frac{\tau+1}{2})}]\\
\end{cases},
\end{equation}
with $c$ an arbitrary normalization constant which is not fixed by the calculation procedure. In the bottom-up approach of modular model building, there is no prescription for the value of $c$; although conventions have been proposed \cite{petcov2023normalisation} in order to set a common ground to compare models in which these normalizations do not play a meaningful role and despite the discussion on phases in \cite{Novichkov_2019} in the context of gCP-invariant modular flavour models, arbitrariness on $c$'s absolute value still remains. As mentioned in \cite{Meloni2023}, under certain circumstances $c$ can be factored out as an overall coefficient and therefore does not contribute in shaping the internal structure of mass matrices; this can easily happen when groups with a triplet $\textbf{3}$ are employed. $S_3$ is not such a group, and we will follow the same strategy outlined in \cite{Meloni2023} to select a specific value for $c$ for our purposes.\\

As pointed out in \cite{Novichkov_2019fundamentaldomain}, thanks to modular symmetry, values of $\tau$ can be restricted to the so-called fundamental domain:
\begin{equation}
    \mathcal{D}=\{\tau\in\mathbb{C}:\quad \text{Im}(\tau)>0,\quad |\text{Re}(\tau)|\le\frac{1}{2},\quad |\tau|>1\}.
\end{equation}
The q-expansions for $Y_1(\tau)$, $Y_2(\tau)$ are:
\begin{equation}
\begin{cases}
    Y_1(\tau)=-c\frac{i\pi}{4}(1 + 24q + 24q^2 + 96q^3+...)\\
    Y_2(\tau)=-ci\pi2\sqrt{3}q^{1/2}(1 + 4q + 6q^2+...)
\end{cases}
\end{equation}
It can be seen from these expressions that the quantity
\begin{equation}
\label{eq:epsilon}
\epsilon\equiv\frac{|Y_2(\tau)|}{|Y_1(\tau)|}\sim |q|^{\frac{1}{2}},
\end{equation}
being $q$ the one in \ref{eq:etadedekind}, is (almost) always exponentially suppressed when $\tau$ gets a vev in its fundamental domain (for $\text{Im}(\tau)>0$). This perturbative parameter (together with the normalization $c$) will be the building block in the construction of models considered in the following.\\
The two modular forms in (\ref{eq:modularforms}) transform as an $S_3$ doublet which, adopting the convention in \cite{Meloni2023}, will be denoted as:
\begin{equation}\label{eq:defY21}
    Y_{\textbf{2}}^{(1)}\equiv \begin{bmatrix}
    Y_1(\tau)\\
    Y_2(\tau)
\end{bmatrix},
\end{equation}
where the lower index labels the representation ($\textbf{1}$ for the singlet, $\textbf{1}'$ for the pseudo-singlet and $\textbf{2}$ for the doublet) and the upper one is the weight $k/2$.\\
All higher-weighted modular forms can be derived by making use of $S_3$ tensor product rules (which are presented in Appendix \ref{app:s3}). It is possible to build at most three linearly indipendent modular forms of weight 4; they are
\begin{equation}
    Y_{\textbf{1}}^{(2)}\equiv Y_1^2(\tau)+Y_2^2(\tau),\quad\quad      Y_{\textbf{2}}^{(2)}\equiv \begin{bmatrix}
    Y_2^2(\tau)-Y_1^2(\tau)\\
    2Y_1(\tau)Y_2(\tau)
\end{bmatrix}
\end{equation}
and the pseudo-singlet combination $\textbf{1}'$ vanishes. Moreover, there are four weight 6 modular forms
\begin{equation}
    Y_{\textbf{1}}^{(3)}\equiv 3Y_1Y_2^2-Y_1^3,\quad Y_{\textbf{1}'}^{(3)}\equiv Y_2^3-3Y_2Y_1^2,\quad Y_{\textbf{2}}^{(3)}\equiv \begin{bmatrix}
Y_1(Y_1^2+Y_2^2)\\
Y_2(Y_1^2+Y_2^2)
\end{bmatrix}.
\end{equation}
Finally, the highest-weighted modular form needed in this discussion will be the weight 8 singlet:
\begin{equation}
    Y_{\textbf{1}}^{(4)}\equiv (Y_1^2+Y_2^2)^{2}.\end{equation}

\section{Building strategy}
\label{strategy}
In this section we explain the procedure we followed to build our model; in particular, we focus  on the origin of hierarchies, since we prefer not to impose them by hand on superpotential coefficients. We first present, as an example, a simplified model for two generations; we then explain how to move to the $3\times3$ case.

\subsection{A simplified \texorpdfstring{$2\times2$}{2x2} case}

Let us start by considering a simplified model with two generations, involving $2\times2$ mass matrices only. The choice of field representations is reported in Table \ref{tab:simplifiedrep}. The two generations of $SU(2)_L$ doublets $l_1$ and $l_2$ are grouped in a doublet $l$ of $S_3$:
\begin{equation}
    l\equiv \begin{bmatrix}
        l_1\\
        l_2
    \end{bmatrix}.
\end{equation}
Up generations of $SU(2)_L$ singlets $u^c_i$ and down generations of $SU(2)_L$ singlets $d^c_i$ are assigned $S_3$ singlet or pseudo-singlet representations, Higgs doublets $H_u$ and $H_d$ are singlets under $S_3$. Each field has a modular charge $k_I$, as remarked in Table \ref{tab:simplifiedrep}.
\begin{table}[h!]
\centering
\begin{tabular}{|c|c|c|c|c|c|c|c|}
\hline
   \rule[-3mm]{0mm}{0.8cm}
   & $l$ & $u_1^c$ & $u_2^c$ & $d_1^c$ & $d_2^c$ & $H_u$ & $H_d$ \\
   \hline
   \rule[-3mm]{0mm}{0.8cm}
   $\Gamma_2\cong S_3$ & $\textbf{2}$ & $\textbf{1}$ & $\textbf{1}'$ & $\textbf{1}$ & $\textbf{1}'$ & $\textbf{1}$ & $\textbf{1}$\\
   \hline
    \rule[-3mm]{0mm}{0.8cm}
   $k_I$ & $k_l$ & $k_{u_1}$ & $k_{u_2}$ & $k_{d_1}$ & $k_{d_2}$ & 0 & 0\\
   \hline
\end{tabular}
\caption{\it Field representations and modular charges for the model.  $l$ is the $SU(2)_L$ "left-handed" doublet containing two generations, $u_i^c$ and $d_i^c$ "right handed" fields, $H_u$ and $H_d$ Higgs doublets.}
\label{tab:simplifiedrep}
\end{table}\\
In order to naturally introduce hierarchies between entries of mass matrices, a possible choice for the values of sums of modular charges for up quarks is:
\begin{equation}
\begin{cases}
    k_{u_1}+k_l=2\\
    k_{u_2}+k_l=6
\end{cases},
\end{equation}
while for down quarks is:
\begin{equation}
\begin{cases}
    k_{d_1}+k_l=2\\
    k_{d_2}+k_l=4
\end{cases}.
\end{equation}
The corresponding superpotential in the up sector is given by:
\begin{equation}
W_u=\; a_uu_1^c(lY_{\textbf{2}}^{(1)})_{\textbf{1}}H_u+b_uu_2^c(lY_{\textbf{2}}^{(3)})_{\textbf{1}'}H_u
\end{equation}
and in the down sector assumes the form:
\begin{equation}
W_d= -\biggl( a_dd_1^c(lY_{\textbf{2}}^{(1)})_{\textbf{1}}H_d+b_dd_2^c(lY_{\textbf{2}}^{(2)})_{\textbf{1}'}H_d\biggl).
\end{equation}
The lower indexes outside parentheses denote the contraction among $l$ and the modular form. We consider a gCP-symmetric theory, hence superpotential coefficients are real. From the superpotential we can derive the mass matrices $M_u$ and $M_d$:
\begin{equation}
M_{u}=v_u
\begin{bmatrix}
a_uY_1 & a_uY_2\\
b_u(Y_\textbf{2}^{(3)})_2 & -b_u(Y_\textbf{2}^{(3)})_1
\end{bmatrix},
\end{equation}
\begin{equation}
M_{d}=v_d
\begin{bmatrix}
a_dY_1 & a_dY_2\\
b_d(Y_\textbf{2}^{(2)})_2 & -b_d(Y_\textbf{2}^{(2)})_1
\end{bmatrix},
\end{equation}
where the lower index outside parentheses labels the doublet component.\\
We can now make some considerations. Quark mass eigenvalues have a hierarchical ordering, and the mixing matrix exhibits hierarchies between different entries as well. We aim to reproduce these features not by imposing hierarchies directly on coefficients in the superpotential, but following the idea in \cite{Meloni2023}: modular charges will play a role similar to Froggatt-Nielsen charges, but without making use of additional flavons. Given the fact that, as seen in the previous section, modular forms of higher weights are constructed via appropriate tensor products starting from modular forms of lower weights, they carry higher powers of the normalization $c$. Once modular charges are assigned to quark fields, the weights of the modular forms that can couple to them are uniquely determined. In this way, the hierarchical structure of the quark sector is encoded in the representation content of the model itself: different entries of the mass matrices are populated by operators carrying different powers of $c$. We can associate a perturbative parameter $\lambda$ to the absolute value of the lowest weight modular forms (which is determined by the choice of $c$); together with the suppression associated with the parameter $\epsilon$ introduced in Eq.~(\ref{eq:epsilon}), this mechanism generates the observed hierarchies while allowing the ratios among the free coefficients to remain naturally of order one.
\footnote{We stress here that the kinetic terms for matter fields in the K\"ahler potential (\ref{eq:kahlerpotential}) are not canonically normalized. In order to get them in the canonical form, we must redefine the fields such that $\varphi^{(I)}\rightarrow[2\text{Im}(\tau)]^{(k_I/2)}\varphi^{(I)}$; hence, a term in the superpotential is multiplied by a factor $[2\text{Im}(\tau)]^{(k/2)}$, where $k$ is the weight of the modular form in the superpotential term. However, the normalization constant $c$ appears in the term with power $k/2$ as well. Therefore, this redefinition can be absorbed in the definition of $c$.}\\
We define
\begin{equation}\label{eq:definizioneYepsilonlambda}
     Y_1\equiv|Y_1|e^{i\phi_1},\quad Y_2\equiv|Y_2|e^{i\phi_2},\quad\quad\quad\quad \epsilon\equiv \frac{|Y_2|}{|Y_1|},\quad \lambda\equiv\frac{1}{|Y_1|},
\end{equation}
where each modular form $Y_i$ is splitted in its absolute value $|Y_i|$ and its phase $\phi_i$. For simplicity, we further define
\begin{equation}
 K_1\equiv 1+e^{-2i(\phi_1-\phi_2)}\epsilon^2,\quad\quad K_2\equiv 1-e^{-2i(\phi_1-\phi_2)}\epsilon^2.
\end{equation}
We can now rewrite mass matrices as
\begin{equation}
M_{u}=v_u|Y_1|^3
    \begin{bmatrix}
        a_ue^{i\phi_1}\lambda^2 & a_ue^{i\phi_2}\lambda^2\epsilon\\
        b_ue^{i(2\phi_1+\phi_2)}K_1\epsilon &  -b_ue^{3i\phi_1}K_1
    \end{bmatrix}
\end{equation}
and
\begin{equation}
M_d=v_d|Y_1|^2
    \begin{bmatrix}
         a_de^{i\phi_1}\lambda &  a_de^{i\phi_2}\lambda\epsilon\\
         2b_de^{i(\phi_1+\phi_2)}\epsilon & b_de^{2i\phi_1}K_2
    \end{bmatrix}.
\end{equation}
Working in the R-L convention, the unitary matrix $L_u$ which contributes to mixing can be determined diagonalizing $M^{\dagger}_uM_u$:
\begin{equation}
    M^{\dagger}_uM^{(diag)}_u=L_u^{\dagger}M^{\dagger}_uM_uL_u.
\end{equation}
In our case, $M^{\dagger}_uM_u$ is perturbatively diagonalized by
\begin{equation}
L_u\approx
    \begin{bmatrix}
        1 & -e^{i(\phi_1-\phi_2)}\epsilon\\
        e^{-i(\phi_1-\phi_2)}\epsilon & 1
    \end{bmatrix},
\end{equation}
where we only show the main non-vanishing term in each entry, but it is understood that this matrix receives corrections of higher order in $\epsilon$ and $\lambda$ and its unitarity holds perturbatively. The eigenvalues are
\begin{equation}
    m^2_{u1}\approx v^2_u|Y_1|^6a^2_u\lambda^4,\quad m^2_{u2}\approx v^2_u|Y_1|^6b_u^2.
\end{equation}
Following the same procedure, $M^{\dagger}_dM_d$ is diagonalized by
\begin{equation}
L_d\approx
    \begin{bmatrix}
         1 & 2e^{i(\phi_1-\phi_2)}\epsilon\\
        -2e^{-i(\phi_1-\phi_2)}\epsilon & 1
    \end{bmatrix}
\end{equation}
with eigenvalues
\begin{equation}
    m^2_{d1}\approx v^2_d|Y_1|^4a^2_d\lambda^2,\quad m^2_{d2}\approx v^2_d|Y_1|^4b^2_d.
\end{equation}
We succeeded in reproducing hierarchies between mass eigenvalues; these are controlled by the $\lambda$ parameter and scale differently in the up and down sector (as needed for quark masses). Moreover, the diagonalizing matrices are equal to the identity plus $O(\epsilon)$ corrections. When we compute the mixing matrix $V=L_u^{\dagger}L_d$ we get (after a phase redefinition)
\begin{equation}\label{eq:V2x23epsilon}
    V=L_u^\dagger L_d\approx
    \begin{bmatrix}
        1 & 3\epsilon\\
        -3\epsilon & 1
    \end{bmatrix},\quad\quad \text{sin}(\theta_{12})\approx3\epsilon.
\end{equation}
Therefore, the Cabibbo angle is controlled by the $\epsilon$ parameter, which mostly depends on $\text{Im}(\tau)$.\\
From the values of the quark observables reported in Table \ref{tab:tabobs}, we can see that $m_{u1}/m_{u2}\equiv m_u/m_c\sim 0.00193$, $m_{d1}/m_{d2}\equiv m_d/m_s\sim 0.0505$ and $\theta_{12}\sim 0.2274$. To leading order, our perturbative analysis gives $m_{u1}/m_{u2}\sim(a_u/b_u)\lambda^2$ and $m_{d1}/m_{d2}\sim(a_d/b_d)\lambda$; therefore, in order to keep $O(1)$ ratios between free parameters, we might fix $\lambda\sim0.04$. From the expression in (\ref{eq:V2x23epsilon}), we obtain $\epsilon\sim0.07$. These results show that in the $2\times2$ case, $\epsilon$ and $\lambda$ are of the same order of magnitude. We will make the hypothesis that this is the case in the $3\times3$ setting as well.
\\
In the next paragraph, we will explain how to extend this procedure to $3\times3$ matrices.\\

\subsection{Extension to \texorpdfstring{$3\times3$}{3x3} matrices and comments on model building}

In \cite{Ra_in_1998}, a procedure is outlined to identify specific structures of $3\times3$ hermitian matrices, exhibiting hierarchies in their entries such that their eigenvalues are hierarchical and mixing angles in the diagonalizing matrix are hierarchical as well (reflecting quark sector hierarchies). Hence, having in mind the Wolfenstein expression for the CKM, we will use these criteria to build mass matrices $M_k$ ($k=u,d$) for quarks such that $M^\dagger_kM_k$ (for which this discussion holds, since they are hermitian) are perturbatively diagonalized by matrices of the form
\begin{equation}
    L_k=\begin{bmatrix}
        1+O(\epsilon^{\alpha_{11}}\lambda^{\beta_{11}}) & O(\epsilon^{\alpha_{12}}\lambda^{\beta_{12}}) & O(\epsilon^{\alpha_{13}}\lambda^{\beta_{13}})\\
        O(\epsilon^{\alpha_{21}}\lambda^{\beta_{21}}) & 1+O(\epsilon^{\alpha_{22}}\lambda^{\beta_{22}}) & O(\epsilon^{\alpha_{23}}\lambda^{\beta_{23}})\\
        O(\epsilon^{\alpha_{31}}\lambda^{\beta_{31}}) & O(\epsilon^{\alpha_{32}}\lambda^{\beta_{32}}) & 1+O(\epsilon^{\alpha_{33}}\lambda^{\beta_{33}})
    \end{bmatrix}.
\end{equation}
$\epsilon$ and $\lambda$ are the same expansion parameters of the $2\times2$ case; $\epsilon$ is the one in (\ref{eq:epsilon}), $\lambda$ is linked to our choice of the normalization $c$ (as stated in the previous paragraph, we have defined $\lambda\equiv1/|Y_1|$, hence $\lambda\sim1/c$). We write them both explicitly to stress that they have different numerical values, but we hypotesize that they are of the same order of magnitude. $\alpha_{ij}$, $\beta_{ij}$ are appropriate exponents needed to reproduce a hierarchical structure similar to the CKM one; this allows to avoid fine-tuned cancellations between entries when computing the CKM matrix $V_{CKM}\equiv L^\dagger_uL_d$. We expect $L_k$ to be the identity at leading order, with the dominant correction in the $(1,2)$ entry. As shown in the $2\times2$ case, this structure can generate a Cabibbo angle of $O(\epsilon)$. We therefore seek textures with an $O(\epsilon)$ contribution to this entry, corresponding to $\alpha_{12}=1$ and $\beta_{12}=0$.
Let us now focus on model building. We have considered in particular $S_3$ representations such that the first two left-handed generations $l_1$ and $l_2$ are grouped in a doublet (as in the case of the $2\times2$ model before)
\begin{equation}\label{eq:ldoublet}
    l\equiv \begin{bmatrix}
        l_1\\
        l_2
    \end{bmatrix}
\end{equation}
and assigned other fields to singlet or pseudo-singlet representations. Other possibilities have been explored, but these assignments have been found to be the most suitable for this framework. Each field has a modular charge, defined in Table \ref{tab:modw}, and even without further specifying the representation assignments of singlets and pseudo-singlets, it is possible to write down the relations modular charges have to satisfy (if we assume we are dealing with combinations of fields allowing a singlet contraction).
\begin{table}[h!]
\centering
\begin{tabular}{|c|c|c|c|c|c|c|c|c|}
\hline
   \rule[-3mm]{0mm}{0.8cm}
   & $l$ & $l_3$ & $u_1^c$ & $u_2^c$ & $u_3^c$ & $d_1^c$ & $d_2^c$ & $d_3^c$ \\
   \hline
   \rule[-3mm]{0mm}{0.8cm}
   $k_I$ & $k_l$ & $k_{l_3}$ & $k_{u_1}$ & $k_{u_2}$ & $k_{u_3}$ & $k_{d_1}$ & $k_{d_2}$ & $k_{d_3}$\\
   \hline
\end{tabular}
\caption{\it Modular charges for quark fields (charges for Higgs fields are set to zero). $l$ is the $SU(2)_L$ "left-handed" doublet containing the first two generations defined in (\ref{eq:ldoublet}), $l_3$ the third "left-handed" generation, $u_i^c$ and $d_i^c$ "right handed" fields.}
\label{tab:modw}
\end{table}\\
In the superpotential there will be terms of the form:
\begin{equation}
    u^c_i\,l\,Y_{u,i,l}\,H_u,\quad\quad d^c_i\,l\,Y_{d,i,l}\,H_d,
\end{equation}
connecting $u^c_i$ and $d^c_i$ to the doublet $l$ and to a modular form ($Y_{u,i,l}$ and $Y_{d,i,l}$) of appropriate weight, and terms
\begin{equation}
    u^c_i\,l_3\,Y_{u,i,l_3}\,H_u,\quad\quad d^c_i\,l_3\,Y_{d,i,l_3}\,H_d,
\end{equation}
connecting $u^c_i$ and $d^c_i$ to the (singlet or pseudo-singlet) $l_3$ field and to a modular form. We can then write relations constraining the possible choices of modular charges and restricting the range of viable superpotential terms:
\begin{equation}
\label{eq:wu}
\begin{cases}
    k_{u_1}+k_l=\alpha\\
    k_{u_2}+k_l=\beta\\
    k_{u_3}+k_l=\gamma\\
    k_{u_1}+k_{l_3}=\delta+\alpha-\gamma\\
    k_{u_2}+k_{l_3}=\delta+\beta-\gamma\\
    k_{u_3}+k_{l_3}=\delta
\end{cases}
\end{equation}
and
\begin{equation}
\label{eq:wd}
\begin{cases}
    k_{d_1}+k_l=\alpha'\\
    k_{d_2}+k_l=\beta'\\
    k_{d_3}+k_l=\delta'+\gamma-\delta\\
    k_{d_1}+k_{l_3}=\delta+\alpha'-\gamma\\
    k_{d_2}+k_{l_3}=\delta+\beta'-\gamma\\
    k_{d_3}+k_{l_3}=\delta'
\end{cases}.
\end{equation}
$\alpha,\alpha',\beta,\beta',\gamma,\delta,\delta'$ are integer numbers and they define the weights of the modular forms which can be coupled to quarks in order to obtain an invariant superpotential. In the $2\times2$ setting, it is possible to choose the values of these sums independently. As it can be seen, the $3\times3$ case is different and some of these values can not be freely chosen. Once values for $\alpha,\alpha',\beta,\beta',\gamma,\delta,\delta'$ are fixed (and all representations, i.e. specific assignments of singlets and pseudo-singlets, are fixed as well), a model is selected (values for charges alone are not fully specified, only values for the sums in (\ref{eq:wu}) and (\ref{eq:wd})).\\

\section{The model: perturbative analysis}
\label{model}
We will now develop our full $3\times3$ model. Building upon the $2\times2$ example presented in the previous paragraph, we will select a value for $c$ in order to have $1/|Y_1|$ as a small perturbative parameter. Field representations are listed in Table \ref{tab:rep2}).
\\
\begin{table}[h]
\centering
\begin{tabular}{|c|c|c|c|c|c|c|c|c|c|c|}
\hline
   \rule[-3mm]{0mm}{0.8cm}
   & $l$ & $l_3$ & $u_1^c$ & $u_2^c$ & $u_3^c$ & $d_1^c$ & $d_2^c$ & $d_3^c$ & $H_u$ & $H_d$ \\
   \hline
   \rule[-3mm]{0mm}{0.8cm}
   $\Gamma_2\cong S_3$ & $\textbf{2}$ & $\textbf{1}'$ & $\textbf{1}$ & $\textbf{1}'$ & $\textbf{1}'$ & $\textbf{1}$ & $\textbf{1}'$ & $\textbf{1}$ & $\textbf{1}$ & $\textbf{1}$\\
   \hline
\end{tabular}
\caption{\it Field representations for the model}
\label{tab:rep2}
\end{table}\\
\\
A gCP symmetry is understood, i.e. superpotential coefficients are real. Modular charges are listed below, there are combinations equal to 8 at most in the up sector:
\begin{equation}
\begin{cases}
    k_{u_1}+k_l=2\\
    k_{u_2}+k_l=6\\
    k_{u_3}+k_l=6\\
    k_{u_1}+k_{l_3}=4\\
    k_{u_2}+k_{l_3}=8\\
    k_{u_3}+k_{l_3}=8
\end{cases},
\end{equation}
while in the down sector the maximum is 6:
\begin{equation}
\begin{cases}
    k_{d_1}+k_l=2\\
    k_{d_2}+k_l=4\\
    k_{d_3}+k_l=4\\
    k_{d_1}+k_{l_3}=4\\
    k_{d_2}+k_{l_3}=6\\
    k_{d_3}+k_{l_3}=6
\end{cases}.
\end{equation}
For up quarks the superpotential assumes the following form:
\begin{equation}
\begin{split}
W_u=\; & a_uu_1^c(lY_{\textbf{2}}^{(1)})_{\textbf{1}}H_u+b_uu_2^c(lY_{\textbf{2}}^{(3)})_{\textbf{1}'}H_u+c_uu_3^c(lY_{\textbf{2}}^{(3)})_{\textbf{1}'}H_u+\\
& B_uu_2^cl_3Y_{\textbf{1}}^{(4)}H_u+C_uu_3^cl_3Y_{\textbf{1}}^{(4)}H_u,
\end{split}
\end{equation}
meanwhile for down quarks we have
\begin{equation}
\begin{split}
W_d= -\biggl(& a_dd_1^c(lY_{\textbf{2}}^{(1)})_{\textbf{1}}H_d+b_dd_2^c(lY_{\textbf{2}}^{(2)})_{\textbf{1}'}H_d+c_dd_3^c(lY_{\textbf{2}}^{(2)})_{\textbf{1}}H_d+\\
& B_dd_2^cl_3Y_{\textbf{1}}^{(3)}H_d+C_dd_3^cl_3Y_{\textbf{1}'}^{(3)}H_d\biggl)
\end{split}
\end{equation}
and, taking vevs into account, the mass matrices are:
\begin{equation}
M_{u}=v_u
\begin{bmatrix}
a_uY_1 & a_uY_2 & 0\\
b_u(Y_\textbf{2}^{(3)})_2 & -b_u(Y_\textbf{2}^{(3)})_1 & B_uY_\textbf{1}^{(4)}\\
c_u(Y_\textbf{2}^{(3)})_2 & -c_u(Y_\textbf{2}^{(3)})_1 & C_uY_\textbf{1}^{(4)}
\end{bmatrix},
\end{equation}
\begin{equation}
M_{d}=v_d
\begin{bmatrix}
a_dY_1 & a_dY_2 & 0\\
b_d(Y_\textbf{2}^{(2)})_2 & -b_d(Y_\textbf{2}^{(2)})_1 & B_dY_\textbf{1}^{(3)}\\
c_d(Y_\textbf{2}^{(2)})_1 & c_d(Y_\textbf{2}^{(2)})_2 & C_dY_{\textbf{1}'}^{(3)}
\end{bmatrix},
\end{equation}
where, for modular forms, the low index outside parentheses labels the doublet component. There are no terms connecting $u_1^c$ and $l_3$ or $d_1^c$ and $l_3$ since there are no modular pseudo-singlets of weight 4.\\
We consider the same definitions as in (\ref{eq:definizioneYepsilonlambda}). In order to match the correct order of magnitude for all matrix elements, we then make the hypothesis
\begin{equation}\label{eq:cdCd}
    c_d=c'_d\epsilon,\quad C_d=\frac{C'_d}{3\epsilon}
\end{equation}
and $c'_d$, $C'_d$ are supposed to have the same order of magnitude of other free parameters; we need a little amount of fine-tuning to get a viable hierarchical structure in the down mass matrix and it is consequently expected that $c_d$ turns out to be slightly smaller than other superpotential coefficients, while $C_d$ slightly bigger. We further define
\begin{equation}\label{eq:defK3K4}
 \begin{split}
 & K_1\equiv 1+e^{-2i(\phi_1-\phi_2)}\epsilon^2,\quad\quad K_2\equiv 1-e^{-2i(\phi_1-\phi_2)}\epsilon^2,\\
 & K_3\equiv 1-3e^{-2i(\phi_1-\phi_2)}\epsilon^2,\quad
      \,\,\, K_4\equiv 1-\frac{1}{3}e^{-2i(\phi_1-\phi_2)}\epsilon^2,
     \end{split}
\end{equation}
hence, factorizing an overall scale, the mass matrices have now the following forms:
\begin{equation}
M_{u}=v_u|Y_1|^4
    \begin{bmatrix}
        a_ue^{i\phi_1}\lambda^3 & a_ue^{i\phi_2}\lambda^3\epsilon & 0\\
        b_ue^{i(2\phi_1+\phi_2)}K_1\lambda\epsilon &  -b_ue^{3i\phi_1}K_1\lambda & B_ue^{4i\phi_1}K_1^2\\
        c_ue^{i(2\phi_1+\phi_2)}K_1\lambda\epsilon &  -c_ue^{3i\phi_1}K_1\lambda & C_ue^{4i\phi_1}K_1^2
    \end{bmatrix}
\end{equation}
and
\begin{equation}
M_d=v_d|Y_1|^3
    \begin{bmatrix}
         a_de^{i\phi_1}\lambda^2 &  a_de^{i\phi_2}\lambda^2\epsilon & 0\\
         2b_de^{i(\phi_1+\phi_2)}\lambda\epsilon & b_de^{2i\phi_1}K_2\lambda & -B_de^{3i\phi_1}K_3\\
         -c'_de^{2i\phi_1}K_2\lambda\epsilon & 2c'_de^{i(\phi_1+\phi_2)}\lambda\epsilon^2 & -C'_de^{i(2\phi_1+\phi_2)}K_4
    \end{bmatrix}.
\end{equation}
Let us first consider an approximation. If we additionally assume
\begin{equation}\label{eq4.9}
    B_u=B'_u\epsilon,\quad B_d=B'_d\epsilon,
\end{equation}
the perturbative analysis of the model leads to particularly simple, readable results.\\
$M^{\dagger}_uM_{u}$ is perturbatively diagonalized by
\begin{equation}
L_u\approx
    \begin{bmatrix}
        1 & -e^{i(\phi_1-\phi_2)}\epsilon & \frac{c_u}{C_u}e^{i(2\phi_1-\phi_2)}\lambda\epsilon\\
        e^{-i(\phi_1-\phi_2)}\epsilon & 1 & -\frac{c_u}{C_u}e^{i\phi_1}\lambda\\
        O(\epsilon^3) & \frac{c_u}{C_u}e^{-i\phi_1}\lambda & 1
    \end{bmatrix}
\end{equation}
and has eigenvalues 
\begin{equation}\label{eq:upeigenvaluesBuBd=0}
    m^2_{u1}\approx v^2_u|Y_1|^8a^2_u\lambda^6,\quad m^2_{u2}\approx v^2_u|Y_1|^8b_u^2\lambda^2,\quad m^2_{u3}\approx v^2_u|Y_1|^8C^2_u.
\end{equation}
$M^{\dagger}_dM_d$ is diagonalized by
\begin{equation}
L_d\approx
    \begin{bmatrix}
         1 & 2e^{i(\phi_1-\phi_2)}\epsilon & \frac{c'_d}{C'_d}e^{i\phi_2}\lambda\epsilon\\
        -2e^{-i(\phi_1-\phi_2)}\epsilon & 1 & -\frac{b_dB'_d}{C'^2_d}e^{i\phi_1}\lambda\epsilon\\
        -\frac{c'_d}{C'_d}e^{-i\phi_2}\lambda\epsilon & \frac{b_dB'_d}{C'^2_d}e^{-i\phi_1}\lambda\epsilon & 1
    \end{bmatrix}
\end{equation}
with eigenvalues 
\begin{equation}
    m^2_{d1}\approx v^2_d|Y_1|^6a^2_d\lambda^4,\quad m^2_{d2}\approx v^2_d|Y_1|^6b^2_d\lambda^2,\quad m^2_{d3}\approx v^2_d|Y_1|^6C'^2_d.
\end{equation}
So, the CKM matrix (when brought into standard parametrization) reads:
\begin{equation}
    V_{CKM}=L_u^\dagger L_d\approx
    \begin{bmatrix}
        1 & 3\epsilon & \frac{c'_d}{C'_d}e^{2i(\phi_2-\phi_1)}\lambda\epsilon\\
        -3\epsilon & 1 & \frac{c_u}{C_u}\lambda\\
        \biggl(3\frac{c_u}{C_u}-\frac{c'_d}{C'_d}e^{-2i(\phi_2-\phi_1)}\biggl)\lambda\epsilon & -\frac{c_u}{C_u}\lambda & 1
    \end{bmatrix}.
\end{equation}
In this setup, each superpotential parameter (except for $B'_u$ and $B'_d$) is directly connected through a simple relation to one observable quantity only (at leading order). The vev of $\tau$ can be estimated from the size of the Cabibbo angle and of the CP-violating phase (from $\epsilon$ and $(\phi_2-\phi_1)$), $B'_u$ and $B'_d$ do not contribute to leading terms in perturbative expressions. We can see that $\lambda$ sets the scale for mass eigenvalues and for the (2,3) angle, while the (1,2) one has the form $\text{sin}(\theta_{12})\approx3\epsilon$ (as in the $2\times2$ case). The (1,3) angle is $O(\lambda\epsilon)$. From this analysis, we are led to consider a value $\lambda\approx0.02$ in order to reproduce appropriate hierarchies among quark observables. This corresponds to fixing the value of the modular forms normalization to:
\begin{equation}\label{eq:deflambda}
 c=i\frac{200}{\pi}.
\end{equation}
Since $B'_u$ and $B'_d$ only affect higher order terms in the calculation of observables, it has indeed been numerically verified that a good fit exists for $B_u,B_d=0$. It seems reasonable to expect that a fit can therefore be found also if the values of $B_u$ and $B_d$ induce small enough perturbations to this scheme. In the following, we will consider the case in which no ad-hoc adjustments of the magnitude of $B_u$ and $B_d$ are required (we will hence suppose that they have $O(1)$ ratios with the other free parameters).\\
If we remove the assumptions in (\ref{eq4.9}), diagonalizing matrices and mass eigenvalues are modified.\\
In the up sector we have:
\begin{equation}
L_u\approx
    \begin{bmatrix}
        1 & -e^{i(\phi_1-\phi_2)}\epsilon & \biggl(\frac{b_uB_u+c_uC_u}{B_u^2+C_u^2}\biggl)e^{i(2\phi_1-\phi_2)}\lambda\epsilon\\
        e^{-i(\phi_1-\phi_2)}\epsilon & 1 & -\biggl(\frac{b_uB_u+c_uC_u}{B_u^2+C_u^2}\biggl)e^{i\phi_1}\lambda\\
        O(\epsilon^3) & \biggl(\frac{b_uB_u+c_uC_u}{B_u^2+C_u^2}\biggl)e^{-i\phi_1}\lambda & 1
    \end{bmatrix},
\end{equation}
with eigenvalues
\begin{equation}\label{eq:upeigenvaluesdef}
    m^2_{u1}\approx v_u^2|Y_1|^8a^2_u\lambda^6,\quad m^2_{u2}\approx v_u^2|Y_1|^8\frac{(B_uc_u-b_uC_u)^2}{B_u^2+C_u^2}\lambda^2,\quad m^2_{u3}\approx v_u^2|Y_1|^8(B^2_u+C^2_u),
\end{equation}
while in the down sector
\begin{equation}
L_d\approx
    \begin{bmatrix}
        1 & L_{d,12} & L_{d,13}\\
        -L_{d,12}^* & 1 & -\frac{b_dB_d}{B^2_d+C'^2_d}e^{i\phi_1}\lambda\\
        -\frac{c'_d}{C'_d}e^{-i\phi_2}\lambda\epsilon & \frac{b_dB_d}{B^2_d+C'^2_d}e^{-i\phi_1}\lambda & 1
    \end{bmatrix},
\end{equation}
where
\begin{itemize}
\item $L_{d,12}=e^{i(\phi_1+\phi_2)}\biggl(\frac{B_dc'_de^{-2i\phi_1}+2b_dC'_de^{-2i\phi_2}}{b_dC'_d}\biggl)\epsilon$,\\
    \item $L_{d,13}=\biggl(\frac{c'_d}{C'_d}e^{i\phi_2}-e^{i(2\phi_1+\phi_2)}\frac{B_d}{C'_d}\biggl(\frac{B_dc'_de^{-2i\phi_1}+2b_dC'_de^{-2i\phi_2}}{B_d^2+C'^2_d}\biggl)\biggl)\lambda\epsilon$\\
\end{itemize}
and the eigenvalues are
\begin{equation}\label{eq:downeigenvaluesdef}
    m^2_{d1}\approx v_d^2|Y_1|^6a^2_d\lambda^4,\quad m^2_{d2}\approx v_d^2|Y_1|^6\frac{b^2_dC'^2_d}{B^2_d+C'^2_d}\lambda^2,\quad m^2_{d3}\approx v_d^2|Y_1|^6(B^2_d+C'^2_d).
\end{equation}
The CKM matrix becomes
\begin{equation}\label{eq:VCKM}
V_{CKM}\approx
    \begin{bmatrix}
        1 & V_{12} & V_{13}\\
        -V_{12}^* & 1 & V_{23}\\
        V_{31} & -V_{23}^* & 1
    \end{bmatrix}
\end{equation}
where
\begin{itemize}
    \item $V_{12}=e^{-i(\phi_1+\phi_2)}\biggl(\frac{B_dc'_de^{2i\phi_2}+3b_dC'_de^{2i\phi_1}}{b_dC'_d}\biggl)\epsilon$,\\
    \item $V_{13}=e^{-i\phi_2}\biggl(\frac{-3b_dB_de^{2i\phi_1}+c'_dC'_de^{2i\phi_2}}{B^2_d+C'^2_d}\biggl)\lambda\epsilon$,\\
    \item $V_{23}=e^{i\phi_1}\biggl(-\frac{b_dB_d}{B^2_d+C'^2_d}+\frac{b_uB_u+c_uC_u}{B_u^2+C_u^2}\biggl)\lambda$,\\
    \item $V_{31}=e^{-i(2\phi_1+\phi_2)}\biggl(\frac{(B_dc'_de^{2i\phi_1}+3b_dC'_de^{2i\phi_2})(b_uB_u+c_uC_u)-b_dc'_d(B_u^2+C_u^2)e^{2i\phi_1}}{b_dC'_d(B_u^2+C_u^2)}\biggl)\lambda\epsilon$.
\end{itemize}
It is now hard to disentangle the structure of the model; a numerical simulation is mandatory in order to test the validity of this perturbative expansion and to find values for the parameters. We will present the procedure and results of our simulation in the next section.

\section{Numerical analysis}
\label{numerical}

First of all, we factor out two overall mass scales: $(v_u|Y_1|^4C_u)$ for the up mass matrix and $(v_d|Y_1|^3C_d)$ for the down mass matrix. These two scales can always be recovered fitting, for example, the absolute values of the top and bottom masses. We therefore fitted CKM parameters and mass ratios, making use of the following set of dimensionless parameters:
\begin{equation}\label{eq:parset}
    \{\text{Re}(\tau),\text{Im}(\tau),\frac{a_u}{C_u},\frac{b_u}{C_u},\frac{c_u}{C_u},\frac{B_u}{C_u},\frac{a_d}{C_d},\frac{b_d}{C_d},\frac{c_d}{C_d},\frac{B_d}{C_d} \}.
\end{equation}
We will call $O_j(p_i)$ the observables obtained from the the $p_i$ input set of parameters and $\overline{O}_j$ the values we want to reproduce, which are summarized in Table \ref{tab:tabobs}. 
\\
\begin{table}[h]
\centering
\begin{minipage}{\textwidth}
\centering
\begin{tabular}{|c|c|c|c|}
\hline
\rule[-3mm]{0mm}{0.8cm}
Observables & $\mu_i\pm1\sigma$ & Observables & $\mu_i\pm1\sigma$ \\
\hline
\rule[-3mm]{0mm}{0.8cm}
$m_{t}/\text{GeV}$ & $89.213 \pm 2.219$ &  $m_{s}/m_{b}$ & $0.0182 \pm 0.0010$ \\
\hline
\rule[-3mm]{0mm}{0.8cm}
$m_{u}/m_{c}$ & $0.00193 \pm 0.00060$ & $\theta_{12}^{q}$ & $0.2274 \pm 0.0007$ \\
\hline
\rule[-3mm]{0mm}{0.8cm}
$m_{c}/m_{t}$ & $0.00280 \pm 0.00012$ & $\theta_{13}^{q}$ & $0.00349 \pm 0.00013$\\
\hline
\rule[-3mm]{0mm}{0.8cm}
$m_{b}/\text{GeV}$ & $0.965 \pm 0.011$ & $\theta_{23}^{q}$ & $0.0400 \pm 0.0006 $  \\
\hline
\rule[-3mm]{0mm}{0.8cm}
$m_{d}/m_{s}$ & $0.0505 \pm 0.0062$ & $\delta_{CP}^{q}/^{\circ}$ & $69.21 \pm 3.11 $\\
\hline
\end{tabular}
\caption{\label{tab:tabobs}\it  Values of observables $\mu_i$ and their $1\sigma$ uncertainties of quark observables when evolved to the GUT scale as calculated in \cite{Antusch_2013}, with the soft SUSY breaking scale $M_{\text{SUSY}}=10$ TeV, $M_{GUT}\approx10^{16}$ GeV and $\tan\beta=5$; table from \cite{ding2024interplaycorrelationsquarklepton}.}
\end{minipage}
\end{table}
\\
To assess the goodness of fit, we introduce a $\chi^2$ function
\begin{equation}\label{eq:chiquad}
    \chi^2_{O_j}(p_i)\equiv\biggl(\frac{O_j(p_i)-\overline{O}_j}{\sigma_j}\biggl)^2,\quad \chi^2(p_i)\equiv\sum_{j=1}^{N_O}\chi^2_{O_j}(p_i),
\end{equation}
where $\sigma_j$ are the uncertainties in Table \ref{tab:tabobs} and $N_O$ is the number of observables.\\
We used this $\chi^2$ to set an acceptance threshold for points in parameter space. The best-fit point was sought using a Metropolis--Hastings-like algorithm implemented in Mathematica. Starting from randomly generated values of the parameters listed in Eq.~(\ref{eq:parset}), the algorithm evaluated all quark observables using the MPT Mathematica package~\cite{Antusch_2005} and computed the corresponding $\chi^2$. In successive runs, the parameter ranges were narrowed to focus on promising regions and identify points with lower $\chi^2$. During this search for the best-fit point, the variation ranges of the ratios $B_u/C_u$ and $B_d/C_d$ were refined in order to keep their values of $O(1)$.

\subsection{Results}

\begin{figure}[ht!]
    \centering
    \includegraphics[width=0.7\linewidth]{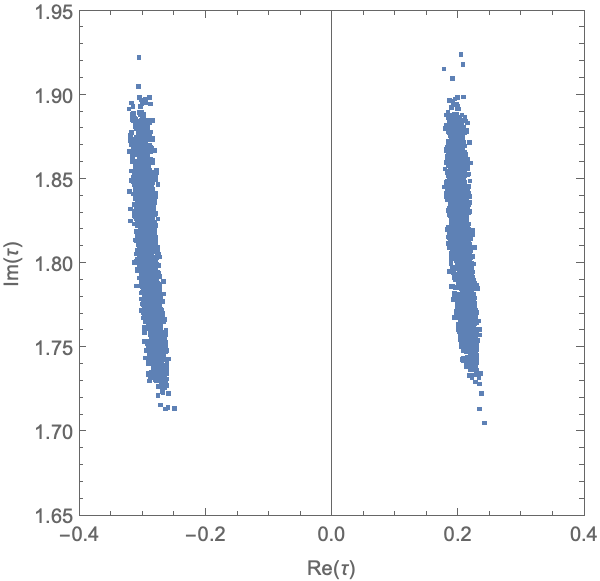}
    \caption{\it Viable regions for $\tau$: there is no reflection symmetry with respect to the values of $\text{Re}(\tau)$. A fit is found both for positive and negative $\text{Re}(\tau)$ due to the interplay with signs of superpotential parameters.}
    \label{fig:ReIm}
\end{figure}

\begin{figure}[ht!]
  \centering
  \begin{subfigure}[b]{0.32\textwidth}
    \centering
    \includegraphics[width=\textwidth]{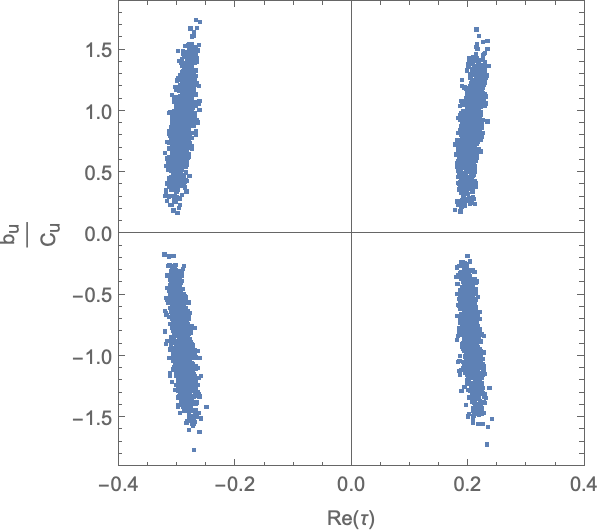}
    \label{fig:subfig1}
  \end{subfigure}
  \hfill
  \begin{subfigure}[b]{0.32\textwidth}
    \centering
    \includegraphics[width=\textwidth]{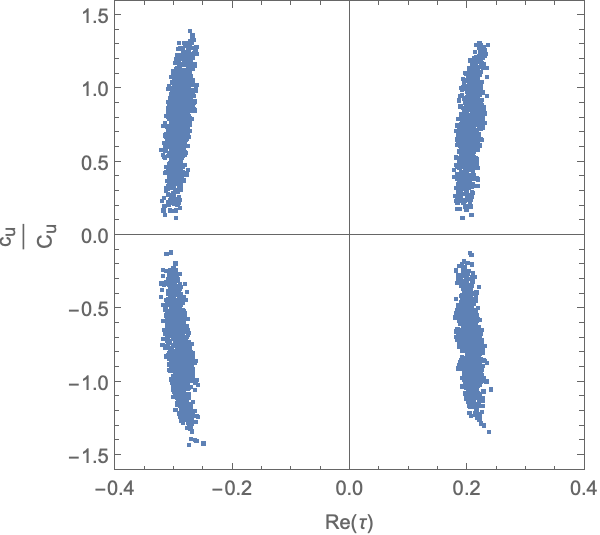}
    \label{fig:subfig2}
  \end{subfigure}
  \hfill
  \begin{subfigure}[b]{0.32\textwidth}
    \centering
    \includegraphics[width=\textwidth]{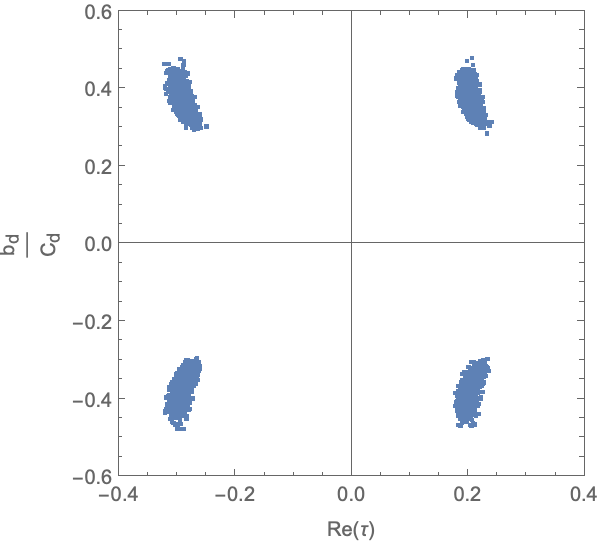}
    \label{fig:subfig3}
  \end{subfigure}
  \hfill
  \begin{subfigure}[b]{0.32\textwidth}
    \centering
    \includegraphics[width=\textwidth]{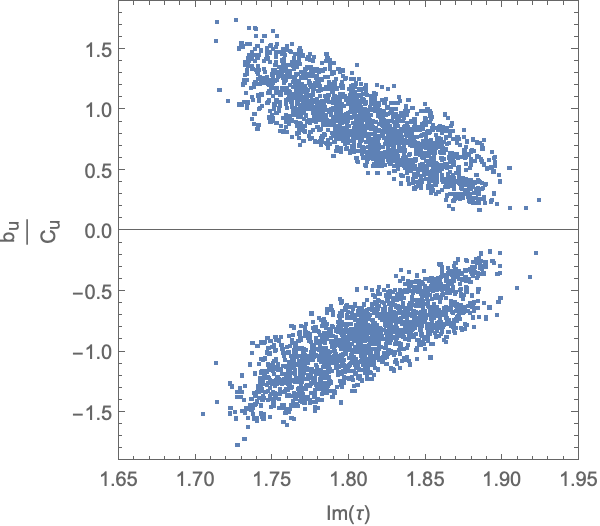}
    \label{fig:subfig4}
  \end{subfigure}
  \hfill
  \begin{subfigure}[b]{0.32\textwidth}
    \centering
    \includegraphics[width=\textwidth]{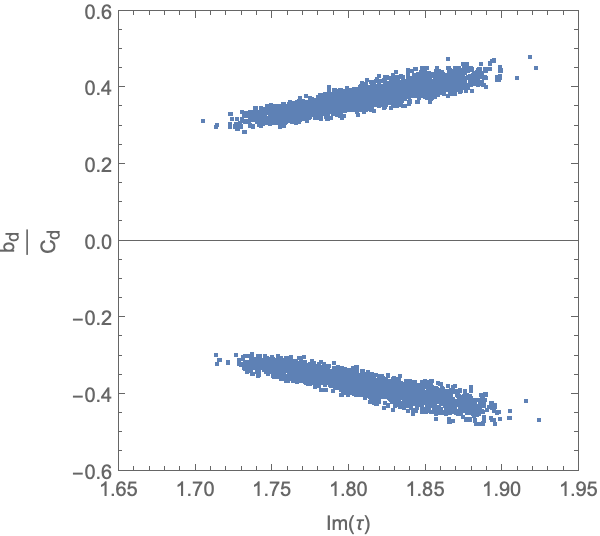}
    \label{fig:subfig6}
  \end{subfigure}
  \hfill
  \begin{subfigure}[t]{0.32\textwidth}
    \centering
    \includegraphics[width=\textwidth]{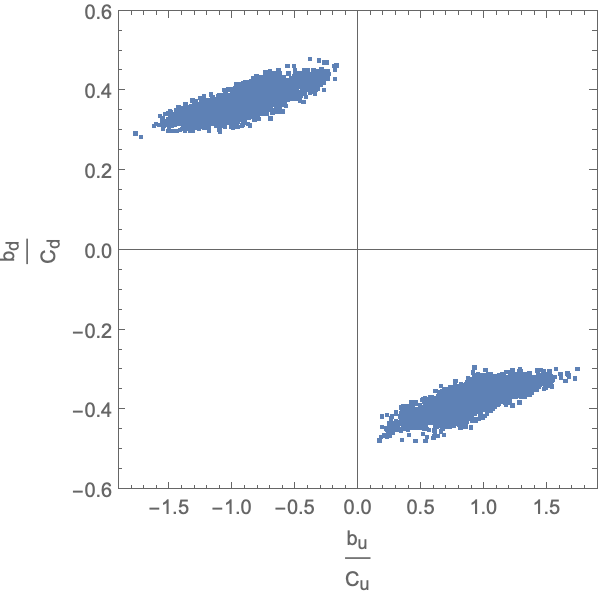}
    \label{fig:subfig8}
  \end{subfigure}
  \caption{\it Some examples of correlations among model parameters. We can see that $\text{Re}(\tau)$, which is responsible for the non-trivial CP phase, is mildly correlated with the parameters $b_u/C_u$, $c_u/C_u$, $b_d/C_d$, which also enter expressions for CKM matrix elements. In the down-right plot we see that $b_u/C_u$ and $b_d/C_d$ are weakly linearly correlated: when one grows, the other grows as well and they have opposite signs. We must remember here that these signs depend on our choice for the fit: $B_u/C_u$, $B_d/C_d>0$; fixing different signs for $B_u/C_u$ and $B_d/C_d$ would also change the signs of $b_u/C_u$ and $b_d/C_d$. The two plots of $\text{Im}(\tau)$ vs $b_u/C_u$ and $b_d/C_d$ show a linear correlation.}
  \label{fig:parpar}
\end{figure}

\begin{figure}[ht!]
  \centering
  \begin{subfigure}[b]{0.45\textwidth}
    \centering
    \includegraphics[width=\textwidth]{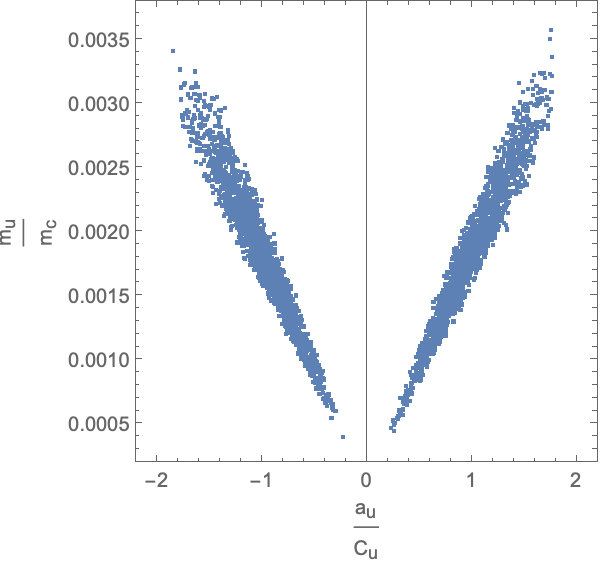}
    \label{fig:subfig25}
  \end{subfigure}
  \hfill
  \begin{subfigure}[b]{0.44\textwidth}
    \centering
    \includegraphics[width=\textwidth]{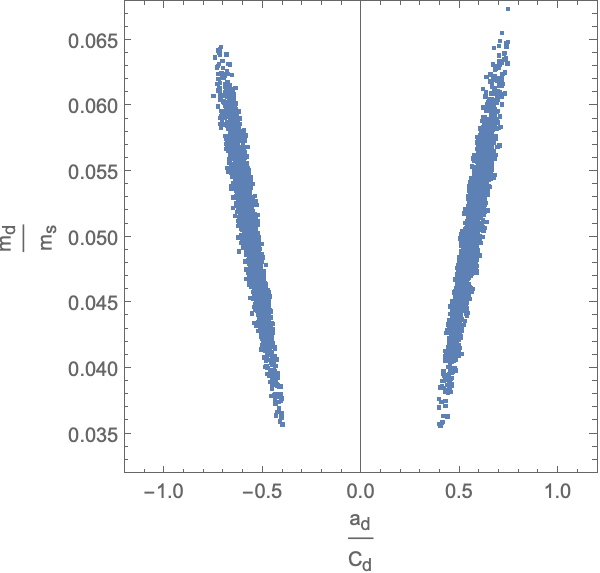}
    \label{fig:subfig26}
  \end{subfigure}
  \caption{\it Correlations between $a_u/C_u$ and $m_u/m_c$ and between $a_d/C_d$ and $m_d/m_s$; these are the only relevant ones among model parameters and observables.}
  \label{fig:parobs}
\end{figure}

\begin{figure}[ht!]
  \centering
  \begin{subfigure}[b]{0.45\textwidth}
    \centering
    \includegraphics[width=\textwidth]{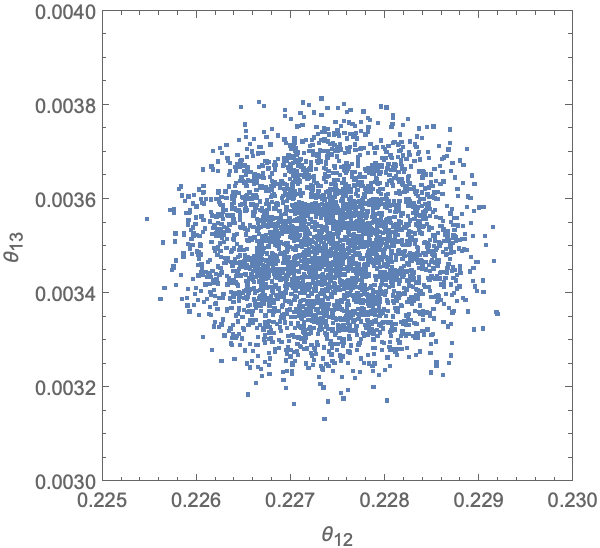}
    \label{fig:subfig31}
  \end{subfigure}
  \hfill
  \begin{subfigure}[b]{0.45\textwidth}
    \centering
    \includegraphics[width=\textwidth]{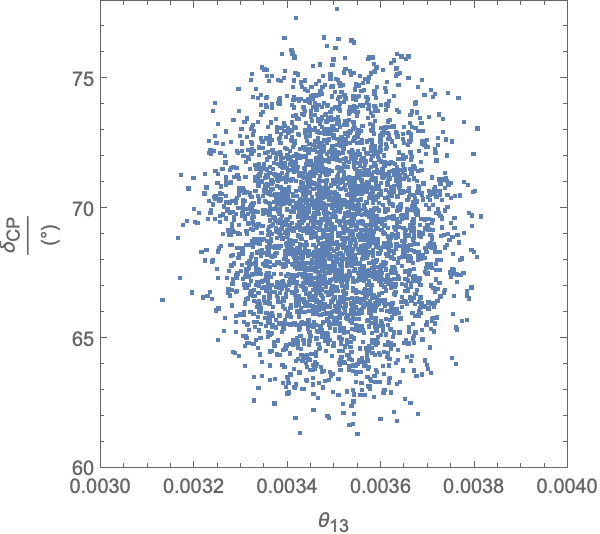}
    \label{fig:subfig32}
  \end{subfigure}
  \caption{\it Examples of correlations among observables: in the left panel we have $\theta_{12}$ vs $\theta_{13}$, in the right panel $\theta_{13}$ vs $\delta_{CP}$. As can be seen, these quantities are not strongly correlated and the same holds for other observables.}
  \label{fig:obsobs}
\end{figure}

We report in Table \ref{tab:fitdefobs} the best-fit values for observables, together with the minimum $\chi^2$ and the individual contributions $\chi^2_{O_j}$ from each observable. As can be seen, data are reproduced with great accuracy. This result is probably a consequence of the number of parameters in the model: we have two more parameters (including $\text{Re}(\tau)$ and $\text{Im}(\tau)$) than observables. This observation is also coherent with the fact that no strong correlations among observables emerge in the model.
\begin{table}[ht]
\centering
\begin{tabular}{|c|c|c|c|}
\hline
    \rule[-3mm]{0mm}{0.8cm}
    $m_u/m_c$ & 0.0019164 & $\chi^2_{m_u/m_c}$ & 0.000513778\\
    \hline
    \rule[-3mm]{0mm}{0.8cm}
    $m_c/m_t$ & 0.00285126 & $\chi^2_{m_c/m_t}$ & 0.182471\\
    \hline
    \rule[-3mm]{0mm}{0.8cm}
    $m_d/m_s$ & 0.0506536 & $\chi^2_{m_d/m_s}$ & 0.000613761\\
    \hline
    \rule[-3mm]{0mm}{0.8cm}
    $m_s/m_b$ & 0.0184897 & $\chi^2_{m_s/m_b}$ & 0.0839261\\
    \hline
    \rule[-3mm]{0mm}{0.8cm}
    $\theta^q_{12}$ & 0.227788 & $\chi^2_{\theta^q_{12}}$ & 0.307523\\
    \hline
    \rule[-3mm]{0mm}{0.8cm}
    $\theta^q_{13}$ & 0.003481 & $\chi^2_{\theta^q_{13}}$ & 0.0047923\\
    \hline
    \rule[-3mm]{0mm}{0.8cm}
    $\theta^q_{23}$ & 0.0401102 & $\chi^2_{\theta^q_{23}}$ & 0.0337639\\
    \hline
    \rule[-3mm]{0mm}{0.8cm}
    $\delta^q_{CP}/^{\circ}$ & 69.3611 & $\chi^2_{\delta^q_{CP}}$ & 0.00236052\\
    \hline
    \multicolumn{1}{|c|}{\rule[-3mm]{0mm}{0.8cm} $\chi^2_{min}$} & \multicolumn{3}{c|}{0.615964}\\
    \hline
\end{tabular}
\caption{\it Best-fit values for the physical observables as obtained from our fitting procedure; also reported are the individual $\chi^2_{O_j}$ and the global $\chi^2_{min}$.}
\label{tab:fitdefobs}
\end{table}\\
\begin{table}[ht!]
\centering
\begin{tabular}{|c|c|}
\hline
\rule[-3mm]{0mm}{0.8cm}
Re($\tau$) & -0.286655\\
\hline
\rule[-3mm]{0mm}{0.8cm}
Im($\tau$) & 1.76807\\
\hline
\rule[-3mm]{0mm}{0.8cm}
    $a_u/b_u$ & -1.07411\\
    \hline
    \rule[-3mm]{0mm}{0.8cm}
    $c_u/b_u$ & 1.18857\\
    \hline
    \rule[-3mm]{0mm}{0.8cm}
    $B_u/b_u$ & 1.17427\\
    \hline
    \rule[-3mm]{0mm}{0.8cm}
    $C_u/b_u$ & 1.04428\\
    \hline
    \rule[-3mm]{0mm}{0.8cm}
    $a_d/c_d$ & -14.9392\\
    \hline
    \rule[-3mm]{0mm}{0.8cm}
    $b_d/c_d$ & -8.68544\\
    \hline
    \rule[-3mm]{0mm}{0.8cm}
    $B_d/c_d$ & 4.78525\\
    \hline
    \rule[-3mm]{0mm}{0.8cm}
    $C_d/c_d$ & 25.6160\\
    \hline
    \rule[-3mm]{0mm}{0.8cm}
    $b_u/c_d$ & 1.49001\\
    \hline
    \end{tabular}
\caption{\it Best-fit values for model parameters.}
\label{tab:fitdefpar}
\end{table}In Table \ref{tab:fitdefpar} best-fit values for parameters are shown. We have rescaled all ratios in (\ref{eq:parset}) dividing by the parameter with the smallest absolute value in each sector in order to describe the best-fit with numbers of absolute value $\ge1$. We present ratios among parameters within the up and down sectors, together with the ratio $b_u/c_d$, which provides the relative normalization between the two sectors. This allows any ratio involving an up-sector and a down-sector parameter to be obtained by appropriately combining the corresponding ratios with $b_u/c_d$.\\
We read that $b_u$ is the smallest parameter in the up sector, where all ratios are very close to 1 and there is substantially no hierarchy among them. $c_d$ is the smallest parameter in the down sector and, as expected, the biggest ratio is between $C_d$ and $c_d$. Being its value $\sim26$, we can say that the model succeeds in reproducing, with a little amount of fine-tuning, the hierarchical ordering of quark observables (which spans five orders of magnitude). Moreover, we can also see from the value of $b_u/c_d\sim1.5$ (which has been derived by reproducing exactly the values of $m_t$ and $m_b$ in Table \ref{tab:tabobs}) that all ratios, even between up and down parameters, are at most $O(10)$.\\
We present in Figures \ref{fig:ReIm}, \ref{fig:parpar}, \ref{fig:parobs} and \ref{fig:obsobs} viable regions for parameters and observables for some cases of interest. These regions have been selected such that points inside them provide $\chi^2\le9$: this ensures that every observable is reproduced inside its $3\sigma$ range. Once $C_u^2$ and $C_d^2$ are factored out, $B_u/C_u$ and $B_d/C_d$ only enter $M_u^\dagger M_u$ and $M_d^\dagger M_d$ squared or multiplied by $b_u/C_u$ and $b_d/C_d$ respectively, as can be seen from the expressions of the mass matrices reported in Appendix \ref{app:matrices}. Therefore, we considered the case $B_u/C_u,B_d/C_d>0$: if a fit is found for certain signs of $b_u/C_u$ and $b_d/C_d$, then a fit exists for other sign combinations preserving the signs of the products $(b_uB_u)/C_u^2$ and $(b_dB_d)/C_d^2$.\\
In Figure \ref{fig:ReIm}, viable regions for the vev of the modulus $\tau$ are shown: solutions have been found both for positive and negative $\text{Re}(\tau)$ and,
as can be seen in Figure \ref{fig:parpar}, solutions with different signs exist for other parameters, too. Some of them exhibit mutual correlations. Looking at the perturbative expressions in (\ref{eq:upeigenvaluesdef}), (\ref{eq:downeigenvaluesdef}) and (\ref{eq:VCKM}), these facts can be explained considering that different model parameters enter in the same expressions for various observables, hence they can balance in magnitude and sign in different ways and still reproduce acceptable values for observables.\\
Given the number of parameters and the structure of the model, when we look for correlations among parameters and observables we find that they are not so evident, for the same reason: there is not, in general, a one-to-one correspondence between one model parameter and an associated observable. This is the case even for the connection between the real part of the modulus $\text{Re}(\tau)$ and the phase $\delta_{CP}$. In the narrow regions of values for $\text{Re}(\tau)$ where a fit is found, there is no strong correlation between it and $\delta_{CP}$. This means that $\text{Re}(\tau)$ is necessary for achieving the CP phase, but the precise value of $\delta_{CP}$ can also be balanced by other parameters (once $\text{Re}(\tau)$ is inside the right region). As we show in Figure \ref{fig:parobs}, strong correlations only emerge between $a_u/C_u$ and $m_u/m_c$ or $a_d/C_d$ and $m_d/m_s$ and it can be understood from our previous perturbative treatment. The parameters $a_u$ and $a_d$ (see eqs. (\ref{eq:upeigenvaluesdef}) and (\ref{eq:downeigenvaluesdef})) appear in the expressions of these mass ratios and do not play a role in leading terms of other observables, so the link between parameter and observable is in these cases stronger.\\
As shown with two examples in Figure \ref{fig:obsobs}, no strong correlations between observables emerge.
\\
\\
\clearpage

\clearpage

\section{Conclusions}
\label{conclusions}

Under the assumption of modular invariance of the action of our theory, we have presented a flavour model for quarks based on the group $\Gamma_2\cong S_3$ and further characterized by a gCP symmetry. Following the strategy in \cite{Meloni2023} and aiming at a minimalistic construction, we introduced no flavons besides the modulus $\tau$ and encoded the origin of hierarchies in the symmetry structure of representations and modular charges.\\
\\
In the mass matrices, hierarchies between entries are governed by the two perturbative parameters $\epsilon$ and $\lambda$ defined in (\ref{eq:definizioneYepsilonlambda}), depending only on the ratio between the two modular forms of lowest weight and their normalization. This choice accounts for the ordering of masses (both in the up and in the down sector) and CKM parameters and implies that free superpotential coefficients have at most $O(10)$ ratios (as can be seen in Table \ref{tab:fitdefpar}). The gCP symmetry leaves the real part of the vev of $\tau$ as the only origin of CP violation and makes all superpotential coefficients real: the model is effectively described by 12 real parameters. Factorizing out two of them (one for up quarks and one for down quarks), 10 parameters are enough to reproduce 8 dimensionless observables (4 mass ratios, 3 CKM mixing angles and the CP phase), all lying in the $1\sigma$ range (as revealed by the numerical analysis, Table \ref{tab:fitdefobs}). This makes our model the $S_3$ modular model for quarks which employs less free parameters.\\
\\
The numerical scan pointed out the existence of appreciable correlations between some of the model parameters (see Figure \ref{fig:parpar}): this means that the values of different parameters can balance in magnitude (and sign) and still provide a good fit for observables. For what concerns correlations between parameters and observables, no noticeable correspondences have been found in general, exception made for the two cases shown in Figure \ref{fig:parobs}. No strong correlations between observables were revealed, as made clear in Figure \ref{fig:obsobs}.

\clearpage

\begin{appendices}

\section{\texorpdfstring{$S_3$}{S3} characters and representations}
\label{app:s3}
\begin{figure}[h]
    \centering
    \includegraphics[width=0.55\linewidth]{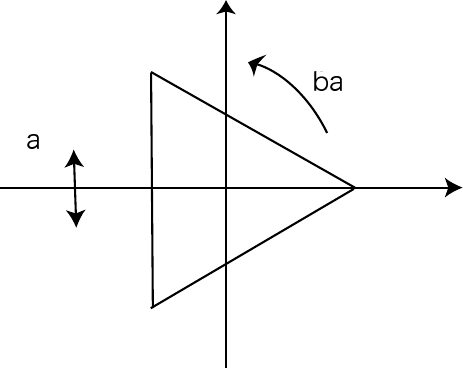}
    \caption{$S_3$ symmetries of an equilateral triangle. Figure from \cite{Ishimori_2010} }
    \label{fig:S3}
\end{figure}

We summarize here some details on $S_3$, following \cite{Ishimori_2010}. $S_3$ is the group of all permutations of three objects $(x_1,x_2,x_3)$ and its order is equal to 6, hence it has six elements corresponding to the following transformations \cite{Ishimori_2010}:
\begin{center}
\begin{equation}
\begin{split}
    e :\quad (x_1,x_2,x_3)\rightarrow (x_1,x_2,x_3),\\
    a_1 :\quad (x_1,x_2,x_3)\rightarrow (x_2,x_1,x_3),\\
    a_2 :\quad (x_1,x_2,x_3)\rightarrow (x_3,x_2,x_1),\\
    a_3 :\quad (x_1,x_2,x_3)\rightarrow(x_1,x_3,x_2),\\
    a_4 :\quad (x_1,x_2,x_3)\rightarrow(x_3,x_1,x_2),\\
    a_5 :\quad (x_1,x_2,x_3)\rightarrow(x_2,x_3,x_1).
    \end{split}
    \end{equation}
\end{center}
$e$ is the identity transformation. Their multiplication forms a closed algebra and the whole group can be generated by two elements $a$ and $b$; thus, by defining $a_1\equiv a$, $a_2\equiv b$ it is possible to write all elements as
\begin{equation}
    \{e,a,b,ab,ba,bab\}.
\end{equation}
The generators satisfy
\begin{equation}
    a^2=b^2=(bab)^2=e,\quad (ab)^3=(ba)^3=e.
\end{equation}
$S_3$ can be also viewed as the group of symmetries of an equilateral triangle, as shown in Figure \ref{fig:S3}; the element $a$ corresponds to a reflection, while $ab$ corresponds to a $(2\pi)/3$ rotation.
\\
\\
As already noticed, this group is characterized by three irreducible representations: a singlet $\textbf{1}$, a pseudo-singlet $\textbf{1}'$ and a doublet $\textbf{2}$, and it has two independent lowest-weighted modular forms of weight $k=2$, whose expression is given in (\ref{eq:modularforms}). In \cite{Kobayashi_2018}, these modular forms have been calculated in the real-symmetric basis of $S_3$, in which the $2\times2$ generators' matrices acting on doublets have the following expression:
\begin{equation}
\rho(a)=
    \begin{bmatrix}
        1 & 0\\
        0 & -1
    \end{bmatrix}, \quad 
    \rho(b)=
    \begin{bmatrix}
        -\frac{1}{2} & -\frac{\sqrt{3}}{2}\\
        -\frac{\sqrt{3}}{2} & \frac{1}{2}
    \end{bmatrix}.
\end{equation}
Meanwhile, the action on pseudo-singlets is
\begin{equation}
\begin{split}
    a:\quad \textbf{1}'\rightarrow-\textbf{1}'\\
    b:\quad \textbf{1}'\rightarrow-\textbf{1}'
    \end{split}
\end{equation}
We now give rules for tensor products. We denote pseudo-singlets with $x$, the components of a first doublet $y_1,y_2$ and of a second doublet $z_1,z_2$. We have the following rules:
\begin{equation}
     \textbf{1}'\otimes\textbf{1}'=\textbf{1}:\quad x_1,x_2\rightarrow x_1x_2,\\
    \end{equation}
    i.e. from two pseudo-singlets we get a singlet,
    \begin{equation}
     \textbf{1}'\otimes\textbf{2}=\textbf{2}:\quad x,\begin{bmatrix}
        y_1\\
        y_2
    \end{bmatrix}\rightarrow \begin{bmatrix}
        -xy_2\\
        xy_1
    \end{bmatrix},
     \end{equation}
     i.e. from a pseudo-singlet and a doublet we get a doublet,
     \begin{equation}\textbf{2}\otimes\textbf{2}=\textbf{1}\oplus\textbf{1}'\oplus\textbf{2}=\begin{cases}
        \textbf{1}\;:\quad \begin{bmatrix}
            y_1\\
            y_2
        \end{bmatrix},\begin{bmatrix}
            z_1\\
            z_2
        \end{bmatrix}\rightarrow y_1z_1+y_2z_2\\
        \\
        \textbf{1}': \quad \begin{bmatrix}
            y_1\\
            y_2
        \end{bmatrix},\begin{bmatrix}
            z_1\\
            z_2
        \end{bmatrix}\rightarrow y_1z_2-y_2z_1\\
        \\
        \textbf{2}\;: \quad \begin{bmatrix}
            y_1\\
            y_2
        \end{bmatrix},\begin{bmatrix}
            z_1\\
            z_2
        \end{bmatrix}\rightarrow \begin{bmatrix}
            y_2z_2-y_1z_1\\
            y_1z_2+y_2z_1
        \end{bmatrix}
    \end{cases},
\end{equation}
i.e. from two doublets we can get a singlet, a pseudo-singlet or a doublet.

\clearpage

\section{Expressions for \texorpdfstring{$M^\dagger_u M_u$}{Mu* Mu} and \texorpdfstring{$M^\dagger_d M_d$}{Md* Md}}\label{app:matrices}

We give here the complete expressions for the hermitian matrices $M^\dagger_u M_u$ and $M^\dagger_d M_d$. In the first case, we write
\begin{equation}
    M^\dagger_u M_u\equiv v_u^2C_u^2\begin{bmatrix}
        m_{11}^u & 
        m_{12}^u &
        m_{13}^u\\
        m_{12}^{u*} &
        m_{22}^u &
        m_{23}^u\\
        m_{13}^{u*} &
        m_{23}^{u*} &
        m_{33}^u
    \end{bmatrix}
\end{equation}
and we have
\begin{equation}
    \begin{split}
        & m_{11}^u\equiv \frac{a_u^2}{C_u^2}|Y_1|^2+\biggl(\frac{b_u^2}{C_u^2}+\frac{c_u^2}{C_u^2}\biggl)|(Y^{(3)}_{\textbf{2}})_2|^2,\\[1.5ex]
      & m_{12}^u\equiv \frac{a_u^2}{C_u^2}Y_1^*\,Y_2-\biggl(\frac{b_u^2}{C_u^2}+\frac{c_u^2}{C_u^2}\biggl)(Y^{(3)}_\textbf{2})_2^*\,(Y^{(3)}_\textbf{2})_1,\\[1.5ex]
      & m_{13}^u\equiv \biggl(\frac{b_u}{C_u}\frac{B_u}{C_u}+\frac{c_u}{C_u}\biggl)(Y^{(3)}_\textbf{2})_2^*\,Y^{(4)}_\textbf{1},\\[1.5ex]
      & m_{22}^u\equiv \frac{a_u^2}{C_u^2}|Y_2|^2+\biggl(\frac{b_u^2}{C_u^2}+\frac{c_u^2}{C_u^2}\biggl)|(Y^{(3)}_\textbf{2})_1|^2,\\[1.5ex]
      & m_{23}^u\equiv -\biggl(\frac{b_u}{C_u}\frac{B_u}{C_u}+\frac{c_u}{C_u}\biggl)(Y^{(3)}_\textbf{2})_1^*\,Y^{(4)}_\textbf{1},\\[1.5ex]
      & m_{33}^u\equiv \biggl(1+\frac{B_u^2}{C_u^2}\biggl)|Y^{(4)}_\textbf{1}|^2.
    \end{split}
\end{equation}
Regarding $M^\dagger_d M_d$, we define
\begin{equation}
    M^\dagger_d M_d\equiv v_d^2C_d^2\begin{bmatrix}
        m^d_{11} & 
       m^d_{12} &
        m^d_{13}\\
        m^{d*}_{12} &
        m^d_{22} &
        m^d_{23}\\
        m^{d*}_{13} &
        m^{d*}_{23} &
        m^d_{33}
    \end{bmatrix},
\end{equation}
where
\begin{equation}
    \begin{split}
        & m^d_{11}\equiv \frac{a_d^2}{C_d^2}|Y_1|^2+\frac{b_d^2}{C_d^2}|(Y^{(2)}_\textbf{2})_2|^2+\frac{c_d^2}{C_d^2}|(Y^{(2)}_\textbf{2})_1|^2,\\[1.5ex]
        & m^d_{12}\equiv \frac{a_d^2}{C_d^2}Y_1^*\,Y_2-\frac{b_d^2}{C_d^2}(Y^{(2)}_\textbf{2})_2^*(Y^{(2)}_\textbf{2})_1+\frac{c_d^2}{C_d^2}(Y^{(2)}_\textbf{2})_1^*(Y^{(2)}_\textbf{2})_2,\\[1.5ex]
        & m^d_{13}\equiv \biggl(\frac{b_d}{C_d}\frac{B_d}{C_d}\biggl)(Y^{(2)}_\textbf{2})_2^*Y^{(3)}_\textbf{1}+\biggl(\frac{c_d}{C_d}\biggl)(Y^{(2)}_\textbf{2})_1^*\,Y^{(3)}_{\textbf{1}'},\\[1.5ex]
        & m^d_{22}\equiv \frac{a^2_d}{C_d^2}|Y_2|^2+\frac{b^2_d}{C^2_d}|(Y^{(2)}_\textbf{2})_1|^2+\frac{c^2_d}{C^2_d}|(Y^{(2)}_\textbf{2})_2|^2,\\[1.5ex]
        & m^d_{23}\equiv -\frac{b_d}{C_d}\frac{B_d}{C_d}(Y^{(2)}_\textbf{2})_1^*Y^{(3)}_\textbf{1}+\frac{c_d}{C_d}(Y^{(2)}_\textbf{2})_2^*Y^{(3)}_{\textbf{1}'},\\[1.5ex]
        & m^d_{33}\equiv |Y^{(3)}_{\textbf{1}'}|^2+\frac{B^2_d}{C^2_d}|Y^{(3)}_\textbf{1}|^2.
        \end{split}
\end{equation}

\end{appendices}

\clearpage

\begingroup
\raggedright
\printbibliography
\endgroup

\end{document}